\documentclass[default]{sn-jnl}

\usepackage[T1]{fontenc} 
\usepackage{xcolor}

\title{Exploring how a Generative AI {\it interprets} music}

\author[1]{Gabriela Barenboim}\email{gabriela.barenboim@uv.es}
\author[2]{Luigi Del Debbio}\email{luigi.del.debbio@ed.ac.uk}
\author[4,5]{Johannes Hirn}\email{johannes.hirn@ext.uv.es}
\author*[1,3]{Ver\'onica Sanz}\email{veronica.sanz@uv.es}
\affil[1]{Departament de F\'{\i}sica Te\`orica and IFIC, Universitat de 
Val\`encia-CSIC, E-46100, Burjassot, Spain}
\affil[2]{University of Edinburgh, UK}
\affil[3]{Department of Physics and Astronomy, University of Sussex, Brighton BN1 9QH, UK}
\affil[4]{Facultad de Ciencias y Tecnolog\unexpanded{í}a, Universidad Isabel I de Castilla, Calle Fern\unexpanded{á}n Gonz\unexpanded{á}lez 76, Burgos, 09003, Spain}
\affil[5]{Centro de Investigaciones Sobre Desertificaci\'on (CIDE),CSIC - Universidad de Valencia - Generalitat Valenciana, 46113 Moncada, Valencia, Spain}

\abstract{
We use Google's MusicVAE, a Variational Auto-Encoder with a 512-dimensional latent space to represent a few bars of music, and organize the latent dimensions according to their relevance in describing music. We find that, on average,  most latent neurons remain silent when fed real music tracks: we call these "noise" neurons. The remaining few dozens of latent neurons that do fire are called "music neurons".

We ask which neurons carry the musical information and what kind of musical information they encode, namely something that can be identified as pitch, rhythm or melody.

We find that most of the information about pitch and rhythm is encoded in the first few music neurons: the neural network has thus constructed a couple of variables that non-linearly encode many human-defined variables used to describe pitch and rhythm. The concept of melody only seems to show up in independent neurons for longer sequences of music.
}

\begin{document} 
\maketitle
\flushbottom

\section{Introduction}

To better understand the manner in which Artificial Intelligence organizes information according to emergent concepts understandable by humans instead of simply memorizing answers thanks to large computing power, we perform several tests on latent neurons in a Variational Auto-Encoder (VAE) designed to reproduce musical melodies.

This is related our previous study~\cite{sym_AI} of symmetry in 2-dimensional images: even though that paper didn't use a VAE, it showed that when we trained Neural Networks (NN) to reproduce contour lines in a given 2D image (equipotentials), the information about the symmetry of the image could still be extracted from the hidden layers of that NN. One limitation of our approach was that a new NN was trained on each new image, and we thus had to employ a statistical procedure (in fact, 2-dimensional PCA and a CNN) to pry the symmetry information from our NN's hidden layers. This in turn implied a loss of information and thus led to imperfect result. An improvement on this procedure would be to replace the ad-hoc dimensional reduction (PCA) by a bottleneck, and to use the whole dataset to train a single NN, instead of training a different NN for each 2D image. The structure of a VAE thus naturally fits our needs.

In fact, some work in this direction has been done by other authors~\cite{angular_momentum}, where VAEs are used to extract the underlying physical invariants (for instance angular momentum) that characterize each example in the distribution. Another example is that of~\cite{Chen:2021twi}, which also uses a VAE to extract the number of independent variables of each mathematical problem in a complex physical system.

As another twist, instead of applying this logic to 2-D potentials, images and then paintings as in~\cite{sym_AI}, we shift to another 2-D discrete problem: music described by discrete pitches in quantized note durations. And instead of different symmetry patterns ---a human-interpretable concept used to summarize and classify potentials and images--- we look for different patterns of rhythms, pitches and melodies in music tracks.

With this in mind, we choose the following VAE as a testing ground: Google Magenta's MusicVAE~\cite{roberts2019hierarchical}, a Variational Auto-Encoder which uses a 512-dimensional latent space to represent a few bars of music.

We ask if the latent vectors are organized in any meaningful pattern, and if these patterns relate to quantities that humans would define when describing the same data, i.e. the concepts of rhythm, pitch and melody.

In Section~\ref{twinkle_section}, we illustrate how MusicVAE works on the first 2 bars of the melody “Twinkle, twinkle, little star”. This allows us to introduce the structure of the latent space, dividing the latent neurons into two set:“ music neurons” and “noise neurons”.

In Section~\ref{latent_section}, we take a large sample of the whole distribution of musical tunes to show that this structure is identical for tracks beyond “Twinkle, twinkle, little star”.

In Section~\ref{jsymb_section} we show which neurons in the latent space encode the information of the human-defined quantities of rhythm and pitch.

In Section~\ref{random_notes_section} we illustrate how sequences of random notes are encoded in the latent space, and use it to test our assumptions about the difference between music neurons and noise neurons.

In Section~\ref{16bar_section} we show that the analysis carries over to chunks of 16 bars of music and discuss the concept of melody.

\section{“Twinkle, twinkle, little star,” and MusicVAE.}
\label{twinkle_section}

For details about the architecture and training procedure of MusicVAE, we refer the reader to the article by the original authors~\cite{roberts2019hierarchical}. For our purposes, it suffices to know that these authors used about 1.5 million MIDI files to create their training dataset, from which those with 4/4 time signature were kept, from which 3.8 million (respectively 11.4 million) monophonic sequences of 2 bars (respectively 16 bars) were extracted.

\begin{figure}[h!]
    \centering
    \includegraphics[scale=0.15]{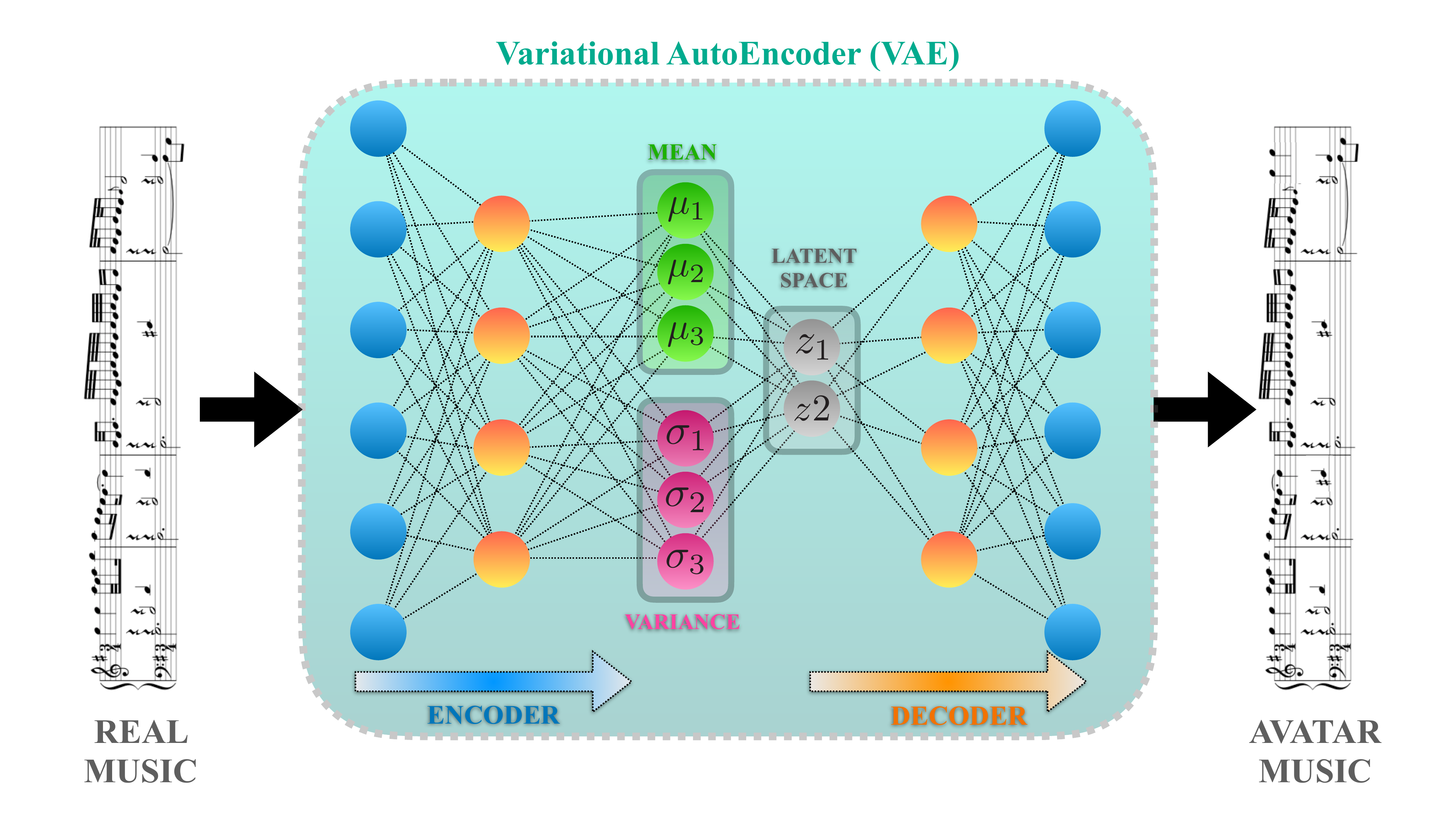}
    \caption{A schematic depiction of the task performed by MusicVAE: a piece of music is encoded into a (multivariate Gaussian) distribution in a so-called 512-dimensional latent space. That distribution can then be sampled from, and decoder back to produce a similar “avatar” piece of music. In reality, the model we use focuses on monophonic melodies and uses Long Short-Term Memory networks in both its encoder and decoder.}
    \label{fig:VAE}
\end{figure}

Any melody used as an input for MusicVAE gets encoded into a vector $\mu_{[1,...,512]}^{\rm twinkle}$ of 512 central values, and a vector $\sigma_{[1,...,512]}^{\rm twinkle}$ of 512 standard deviations, defining a 512-dimensional Gaussian distribution in latent space. See Fig.~\ref{fig:VAE} for a schematic depiction of the process.

\begin{figure}[h!]
\centering 
\includegraphics[width=1\textwidth,
]{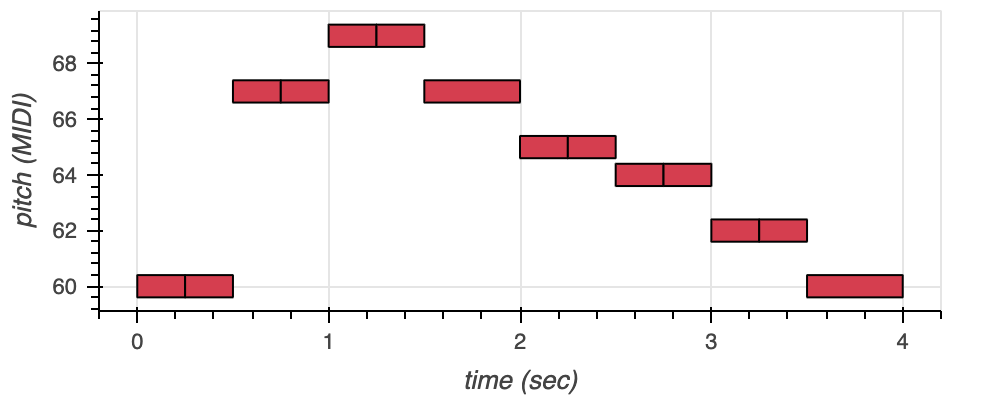}
\vfill
\includegraphics[width=1\textwidth,
]{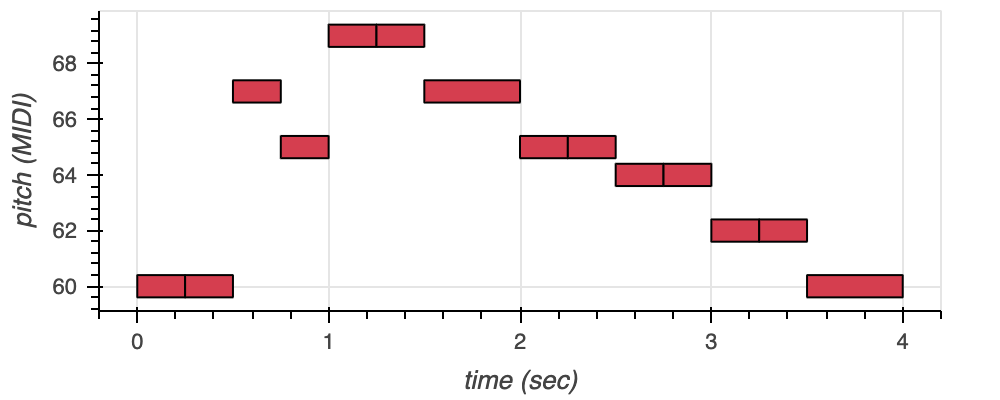}
\caption{\label{twinkle_MIDI} {\em Top:} The first two bars of the melody for “Twinkle, twinkle, little star” with note frequency encoded as pitch in MIDI notation. {\em Bottom:} The result of decoding a random sample from the encoded distribution for “Twinkle, twinkle”.}
\end{figure}

Sampling from this 512-dimensional distribution and passing it through the decoder part of MusicVAE yields back another 2-bar note sequence that is similar, but not identical to the original one, see Figure~\ref{twinkle_MIDI}.

For instance, focusing as an example on the melody “Twinkle, twinkle”, one starts from its sheet music, or rather piano-roll representation as depicted at the top of Fig.~\ref{twinkle_MIDI}, where the $x$-axis depicts time and the $y$-axis depicts frequency encoded in a logarithmic scale (equivalent to numbering the corresponding keys on a piano keyboard from left to right, including black keys). Using this melody as an input and sending it through MusicVAE typically return an output such as the lower part of Figure~\ref{twinkle_MIDI}, i.e. a variation on the melody that still respects the key and rhythm patterns of the original, but is not necessarily identical to the original.

In fact, given the probabilistic nature of the VAE, the output will be slightly different each time we run it on the same input. We illustrate this in the case of our example below in Figure~\ref{twinkle_latent}. The top plot depicts the standard deviations $\sigma_{[0,...,511]}^{\rm twinkle}$ of the Gaussian distribution for each of 100 of the 512 latent dimensions, sorted from smallest $\sigma$ to largest. The middle plot shows the first 100 components of the central value $\mu_{[1,...,512]}^{\rm twinkle}$ in the same order. Finally, the bottom plot represents a sample drawn from the distribution described by the standard deviations and means represented in the top two plots. As most sample, this one approximately reproduces the central values for the first few dozen latent neurons (for the order we have chosen), and then produces random noise for the other ones.

In fact the bottom plot in Figure~\ref{twinkle_MIDI}, when decoded, yields the melody at the bottom of Figure~\ref{twinkle_MIDI}, while decoding the central values themselves (i.e. the values in the middle plot of Figure~\ref{twinkle_latent}) yields back the original version of “Twinkle, twinkle, little star”.

In the top two plots of Figure~\ref{twinkle_latent}, we notice that 475 dimensions have $\sigma \approx 1$ and $\mu \approx 0$, while only 37 dimensions have $\sigma < 1$ and most of these have $\mu$ visibly different from 0. 

The bottom plot in the Figure shows a random sample from the distribution defined by the central values from the middle plot in the same Figure, and the standard deviations from the top of the same Figure. 

\begin{figure}[h!]
\centering 
\includegraphics[width=1\textwidth,
]{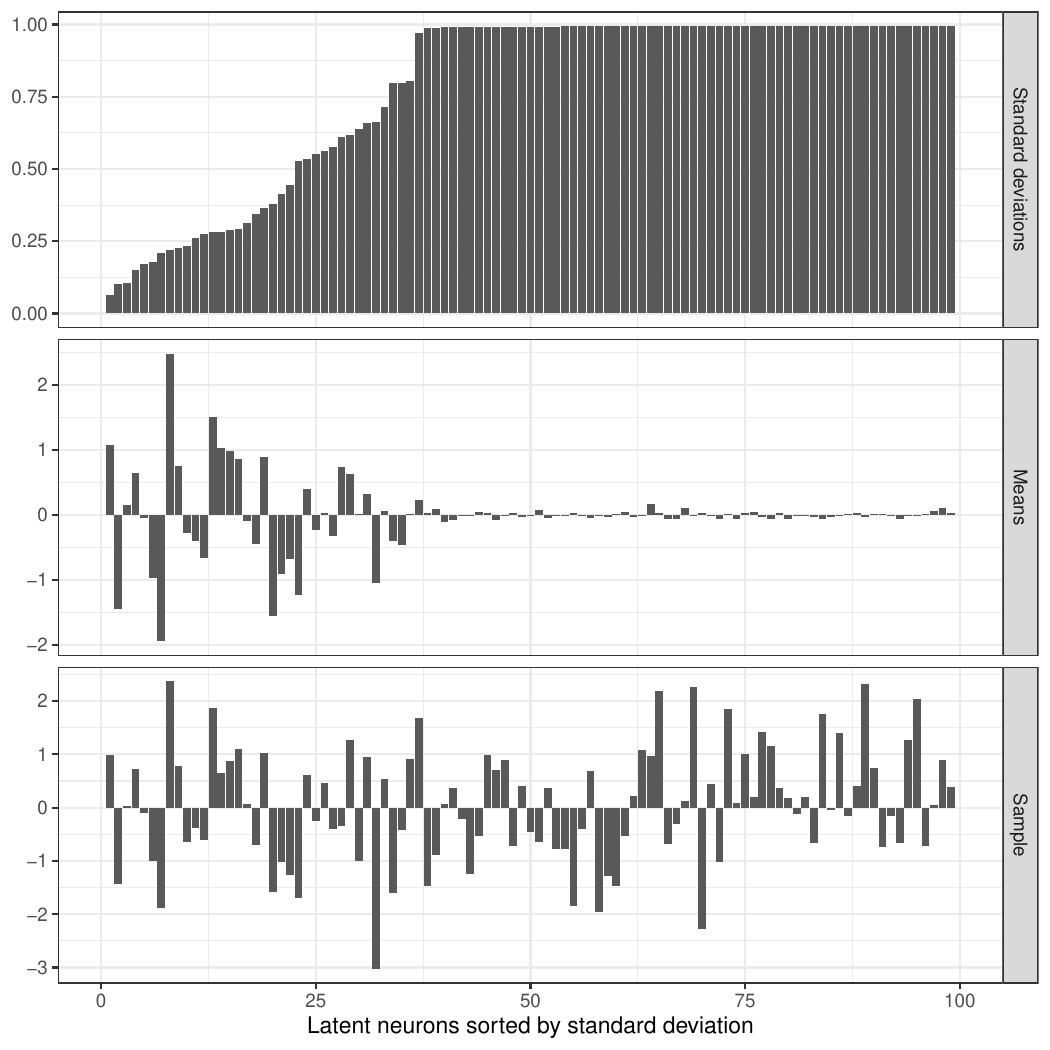}
\caption{\label{twinkle_latent}Latent encoding for the note sequence in Figure~\ref{twinkle_MIDI}. We only show the 100 dimensions with the smallest variance, in order. Standard deviations are at the top, means in the middle, and a sample extracted from the distribution at the bottom.}
\end{figure}

\section{The structure of MusicVAE's latent space}
\label{latent_section}

For the application described here, we used MusicVAE's 2-bar models, which considers monophonic sequences of notes quantized down to 16th notes: this yields 32 discrete time stamps for a note to start in the note sequence. As for the pitches, MIDI files specify $2^7$ discrete pitches, i.e. slightly more than the 88 keys on a piano, to which we need to add the possibility of starting a silence, and the possibility of holding a frequency for the next interval, i.e. 130 discrete possibilities. With this quantization, there are $4 \times 10^{67}$ possible 2-bar note sequences, whereas a latent space of 512 neurons —even if these neurons were perceptrons (i.e. binary neurons)— can describe over $1 \times 10^{154}$ possibilities, and is thus overdimensioned even if  we consider all possible note sequences.
If we now restrict ourselves to the subspace of note sequences that can be called “music”, the dimension of the space is probably even smaller. The question is: how many neurons do we really need in order to describe music?

\begin{figure}[h!]
\centering 
\includegraphics[width=1\textwidth,
]{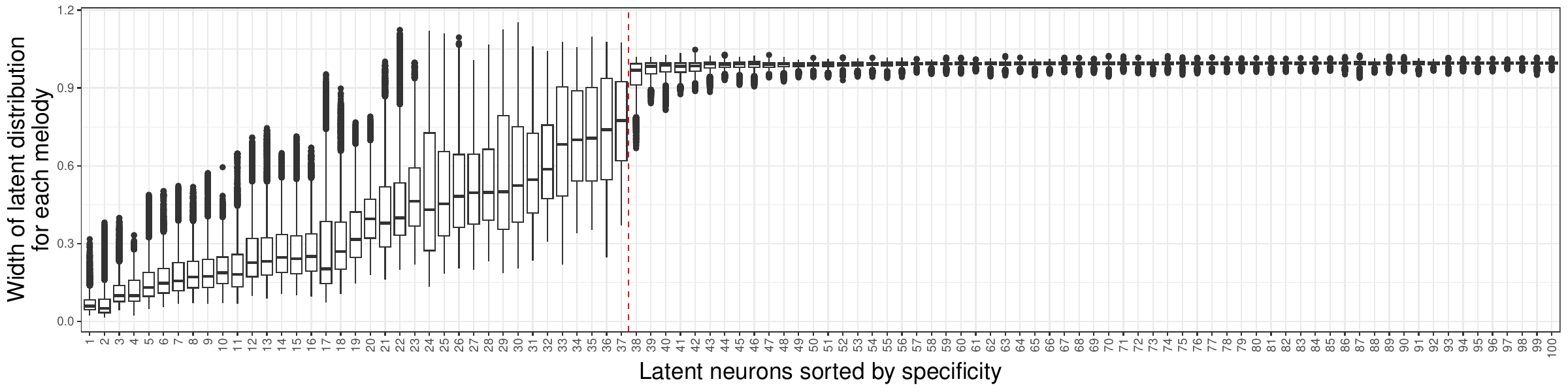}
\includegraphics[width=1\textwidth,
]{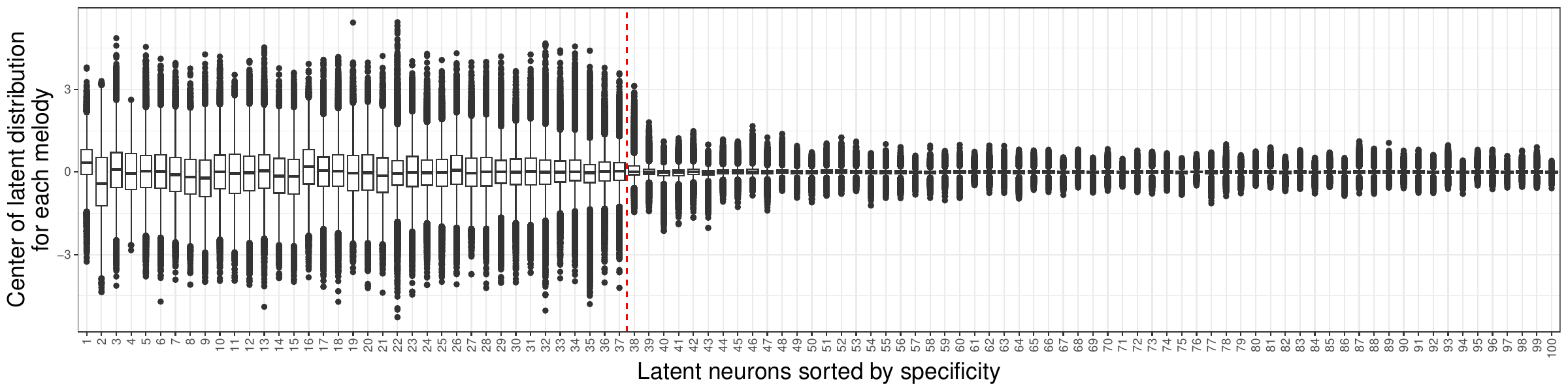}
\caption{\label{latent_2bar} Boxplot of standard deviations  (top) and central values (bottom) for the 2-bar model run on 2,000 random tracks from our dataset.}
\end{figure}

To begin answering this question, we can ask whether the pattern observed in Figure~\ref{twinkle_latent} is true for other pieces of music, i.e. if one latent dimension is specified with great precision (small spread) for one music piece mean, will it be true for other tracks? The answer is a resounding “YES”, as demonstrated by Figure~\ref{latent_2bar}.

To establish this, we pick at random from the 100,000 tracks in the LMD dataset~\cite{lmd}, which is itself contained in the larger dataset that was used to train the model~\cite{roberts2019hierarchical}. We use about 10,000 tracks, from which the standard algorithm in MusicVAE extracts 5 melodies each, yielding 50,000 melodies.
We then encode these melodies, and study their  encodings in latent space.

We can see that each note sequence's encoding is always very precisely specified along the first few dimensions in this ordering. This presumably means that the VAE uses these dimensions to encode a lot of important information about the music, whereas the dimensions with large spread do not contain much information.

We can check this by looking at how much the central values of each melody vary along these same dimensions. In the lower plot of Figure~\ref{latent_2bar}, we depict the central values for a random set of tracks in our database, using the same 100 latent coordinates as the top plot. The first thing to notice is that, for all dimensions with a wide spread (to the right of the plots), the central value is very close to 0: all note sequences are encoded with the same unit Gaussian for these hundreds of dimensions, and hence we have hundreds of dimensions in latent space that carry very little musical information. We call these latent neurons "noise neurons".

On the other hand, for the "music neurons", i.e. the 37 latent neurons to the left of the vertical dashed red line in Figure~\ref{latent_2bar}, which have narrow distributions, we find that the central value fluctuates from track to track: these are the dimensions that contain the actual information about music. Not only are the central values of the music neurons well-specified for each song, they vary a lot from song to song.

\begin{figure}[h!]
\centering 
\includegraphics[width=1\textwidth,
]{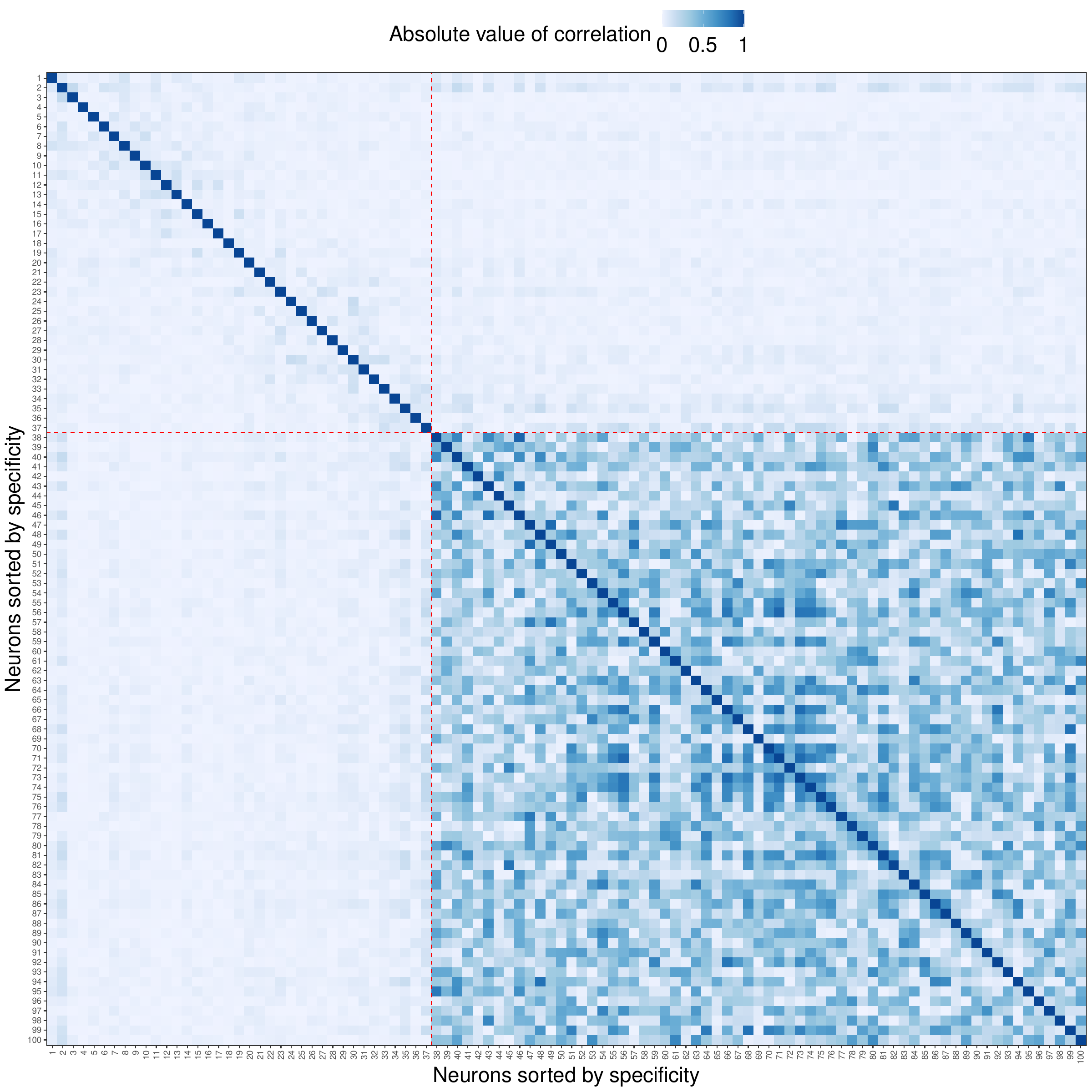}
\caption{\label{latent_corrs_2bar} Pearson correlations of central values for the first 100 latent dimensions, between melodies extracted from a random sample of real music tracks.}
\end{figure}

For our dataset, the correlation matrix in latent space is shown in Figure~\ref{latent_corrs_2bar}: at least, the 37 dimensions we identified in Figure~\ref{latent_2bar} are quite uncorrelated, and this is the subspace we are interested in. Note that these are correlations between central values, not between samples from the distributions, in which case we would also get an approximate unit matrix for the lower-right corner\footnote{This is due to the Kullback-Leibler divergence in the loss function, which regularizes the latent distribution so that it approximates a standard normal prior.}.

\section{Neurons for pitch and rhythm}
\label{jsymb_section}

In this Section, we attempt to disentangle how MusicVAE stores the information about music's most fundamental qualities rhythm and pitch.

We use the {\tt music21} library\footnote{\url{https://web.mit.edu/music21/}
} to extract some human-defined variables that are used to describe music (music features) as listed in the {\tt jSymbolic} manual\footnote{\url{ http://jmir.sourceforge.net/manuals/jSymbolic_manual/home.html}} to quantify the rhythm, pitch and melody of the tracks in our dataset\footnote{To simplify the problem, we only include scalar music features, and not vectors such as histograms.}. To compute correlations that capture non-linear dependencies, we use the correlation coefficient defined in the Python {\tt phik} library~\cite{phik}.

\begin{figure}[h!]
\centering 
\includegraphics[width=1\textwidth,
]{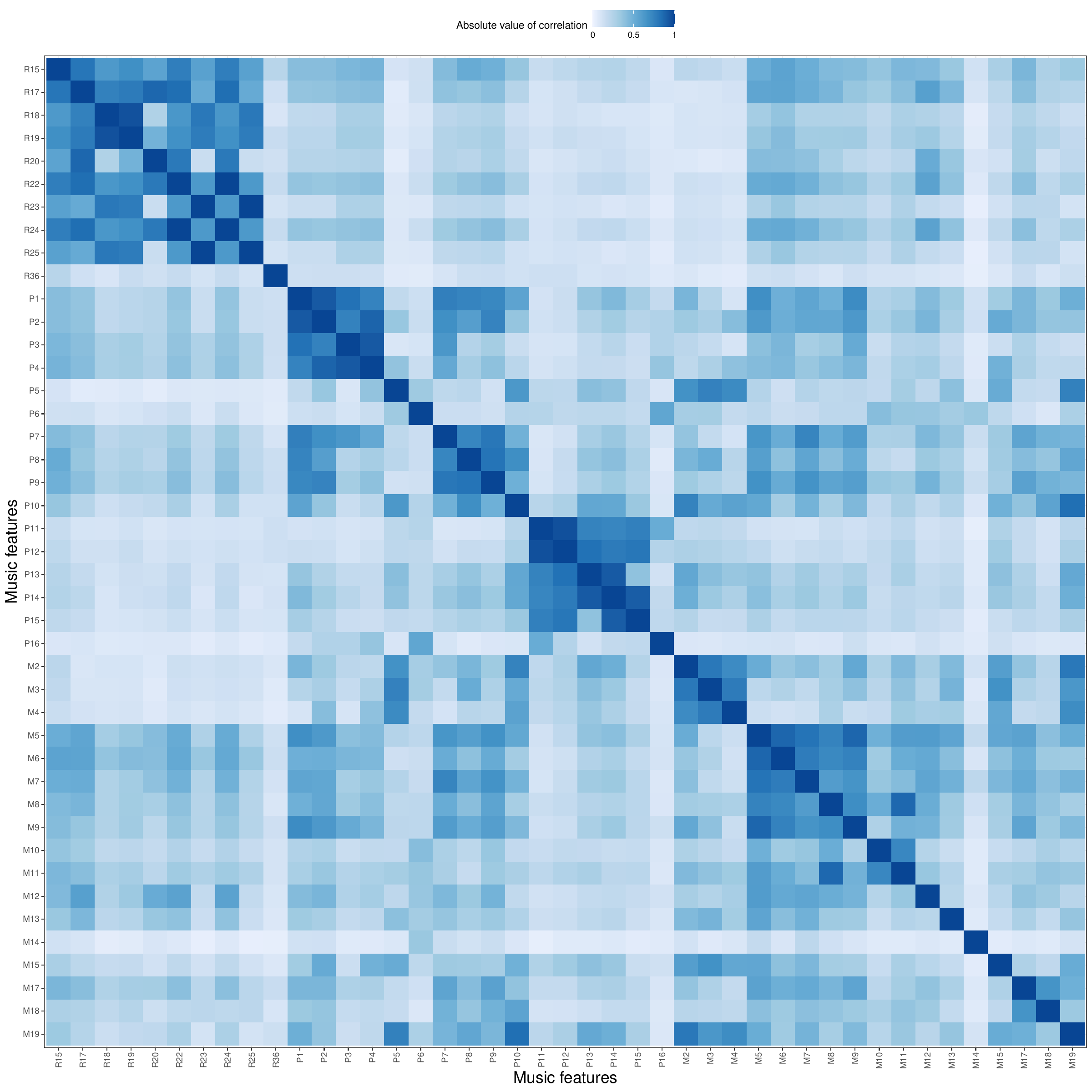}
\caption{\label{fig_jsymb_corrs} Pearson correlations between {\tt jSymbolic} music features in our dataset. The first letter of each feature indicates the type: R for rhythm, P for pitch and M for melody.}
\end{figure}

The result is shown in Figure~\ref{fig_jsymb_corrs}. We can see that rhythm features (starting with the letter R at the bottom right in the Figure) are heavily correlated amongst themselves, as are some subsets of pitch features (from P11 to P15). Other pitch and melody features also form subsets of highly-correlated features, but tend to correlate with other groups as well, even groups starting with a different letter (R or P).

\begin{figure}[h!]
\centering 
\includegraphics[width=.9\textwidth,
]{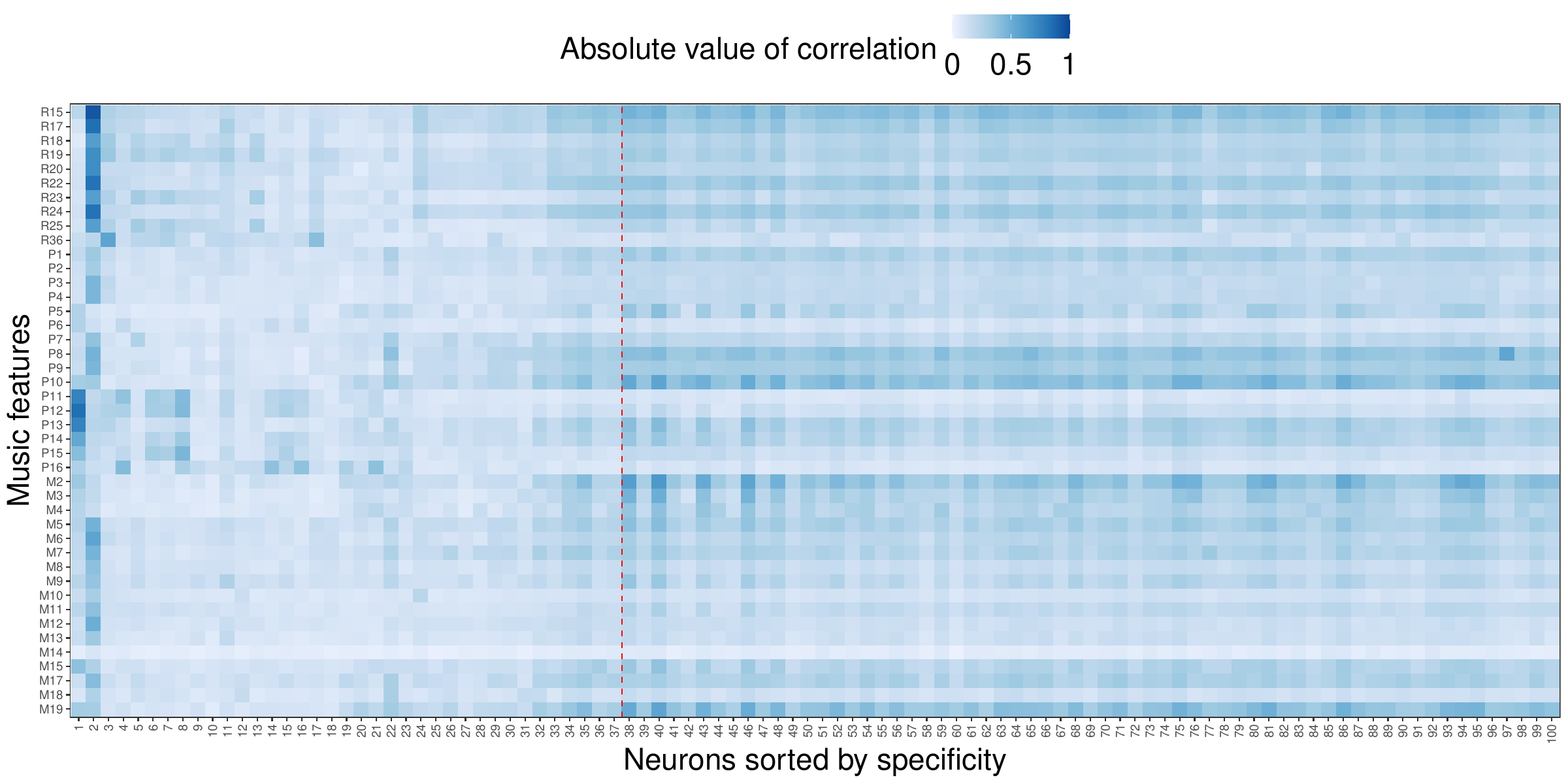}
\caption{\label{fig_jsymb_latent_corrs_2bar} Nonlinear {\tt phik} correlations between {\tt jSymbolic} features and latent neurons, with neurons sorted by the same order as in the previous plots. We can easily pick out the first neuron as being heavily correlated with many rhythm features, and the second one with many pitch features.}
\end{figure}

Within our set of music tracks, we can compute correlations between latent neuron central values and {\tt jSymbolic} music features, as displayed in Figure~\ref{fig_jsymb_latent_corrs_2bar}.

One could follow the standard procedure of attempting to define attribute vectors \cite{roberts2019hierarchical}, by computing the difference in latent space between the averages of subsets of songs with extreme values for a given symbolic feature. We have not found this to be useful here, as the vectors thus obtained actually end up less strongly correlated with symbolic features than do single neurons\footnote{This can seem surprising, as the attribute vectors are defined as linear combinations of latent dimensions, and shouldn't fare worse than individual dimensions. Yet, the averaging procedure introduce some imperfections.} From our point of view, this is one of the important results of this paper: the VAE has used 37 coordinates to represent the musical information in the songs, and among those, it has selected the first two of its canonical dimensions ---as opposed to linear combinations--- to carry most of the information that humans would describe as pitch in the first dimension, and rhythm in the second dimension.

\begin{figure}[h!]
\centering 
\includegraphics[angle=-90,width=1\textwidth,
]{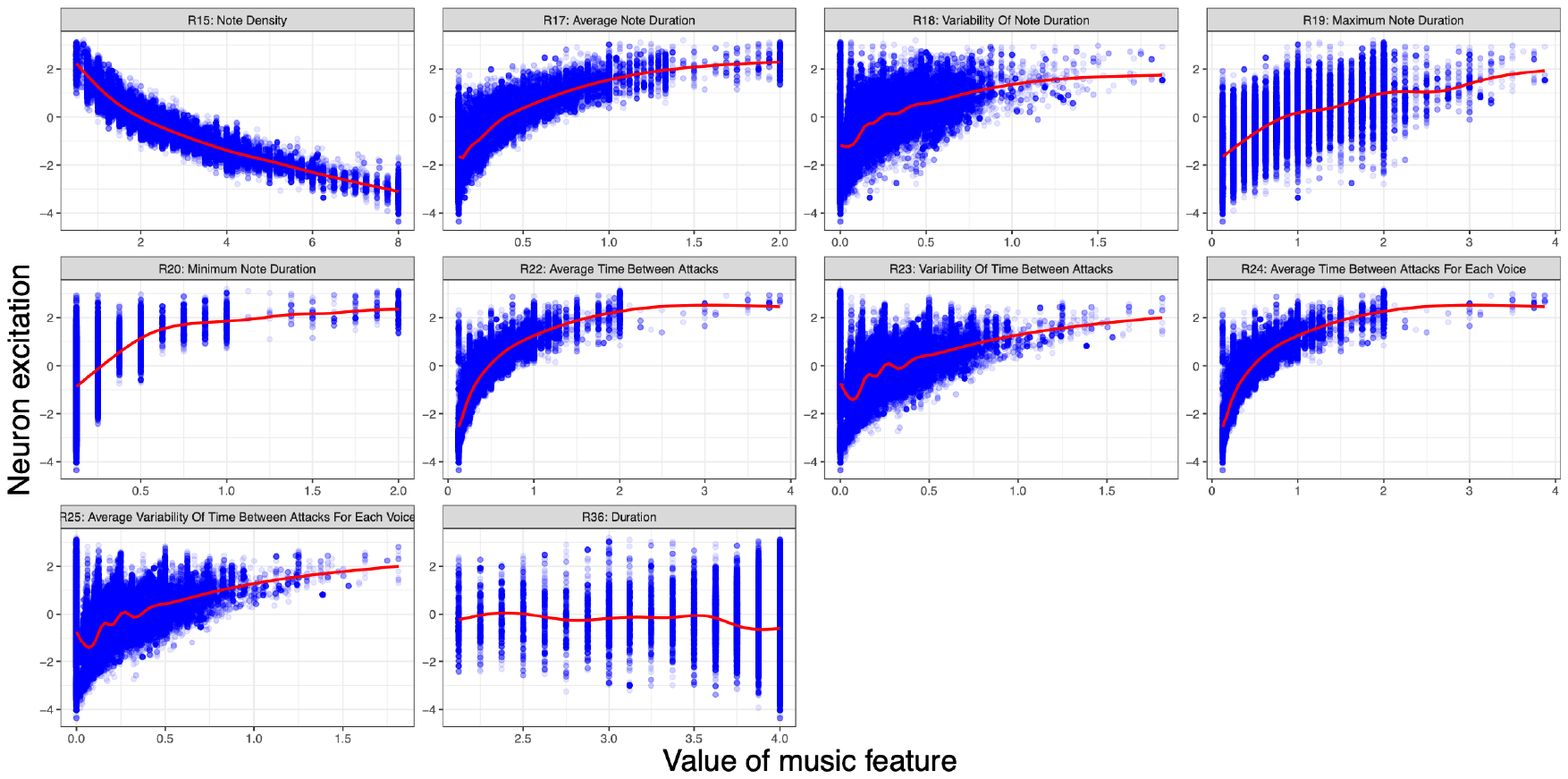}
\caption{\label{R_neuron_R_features}Activation of the second neuron sorted by specificity, against the value of several rhythm features. The red curves show local regression fits of latent neuron excitation as a function of the music feature.}
\end{figure}

Figure~\ref{R_neuron_R_features} shows that the second neuron is highly correlated with some rhythm features.

\begin{figure}[h!]
\centering 
\includegraphics[width=1\textwidth,
]{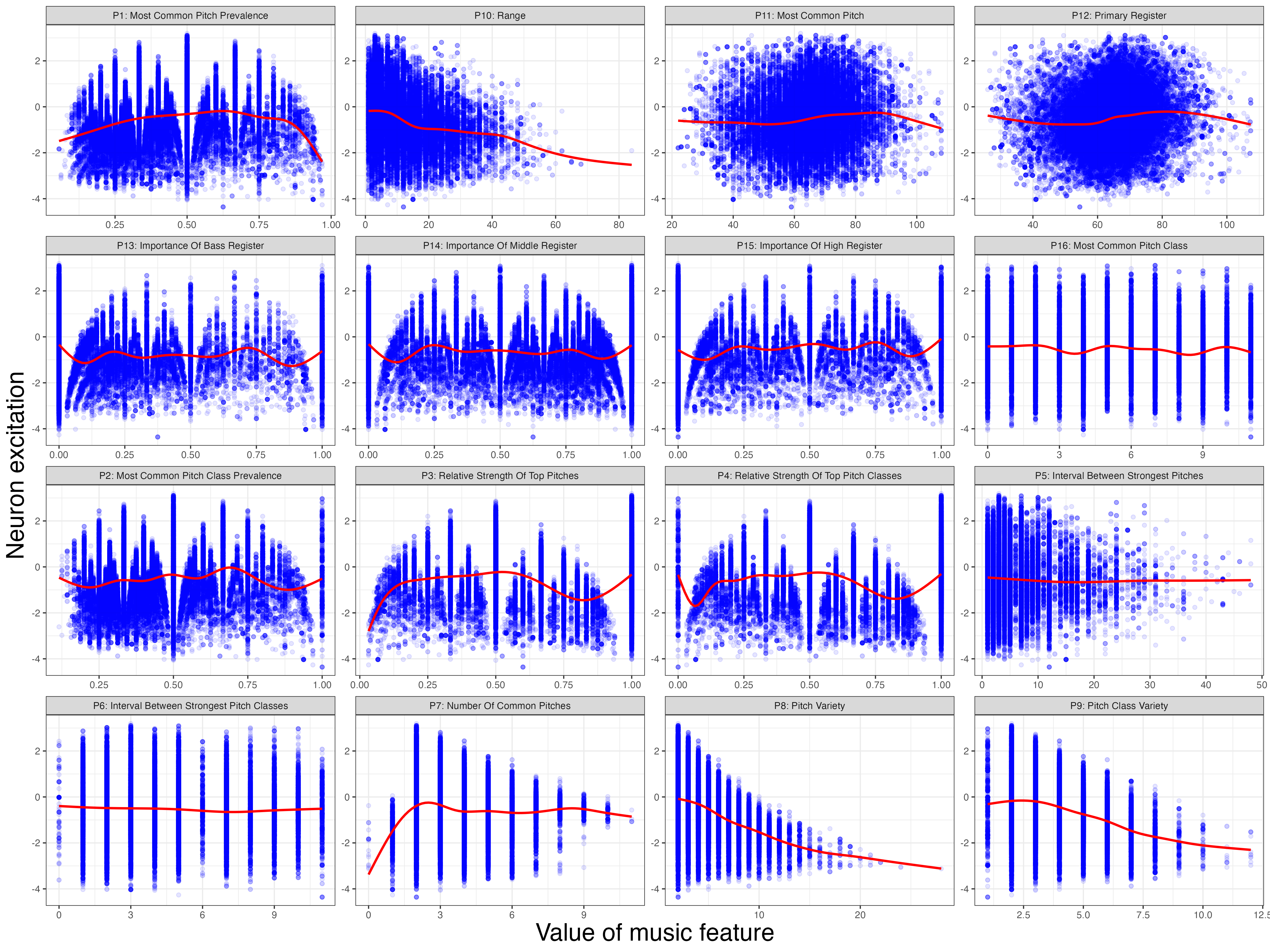}
\caption{\label{R_neuron_P_features}Activation of the second neuron sorted by specificity, against the value of several pitch features. The red curves show local regression fits.}
\end{figure}

In Figure~\ref{R_neuron_P_features}, we can see that this same latent neuron is not correlated with pitch features. 

\section{Random sequences of notes}
\label{random_notes_section}

MusicVAE has been trained on a million tracks of music of various genres, and encodes these tracks and others using essentially the 37 music neurons. We also saw that MusicVAE mostly uses 2 neurons to represent pitch and rhythm. We can now ask what happens if we provide an input that is not real music: where does it get encoded in latent space? Are the 37 music neurons enough to distinguish music from noise, and what about the two main pitch and rhythm neurons?

\subsection{Generating the data}

We create 50,000 note sequences by switching on a random number of notes. We draw the number of note-on events from a uniform distribution on integers from 2 to 32, as 32 sixteenth notes would fill exactly two bars at tempo 4/4. Notes can only be switched on at discrete intervals (every sixteenth note), and we make sure that there is always exactly one note being played at any given time\footnote{To achieve this, each note is then extended from it switch-on time (note-on event) until the start of the next note. To avoid boundary issues, the first note starts at time 0, while the last one extends until the end of the second bar. The total duration of the two bars is set by drawing from a uniform distribution of integer number of seconds between 1 and 8 (both inclusive).}.

The pitch for each note is selected from a uniform distribution over the integers from 30 to 100 in midi notation. This means that the pitches of the various notes in the sequence are uncorrelated, and as such, follow no melodic structure.

\subsection{Distinguishing music and noise using individual neurons}

\begin{figure}[h!]
\centering 
\includegraphics[width=1\textwidth,
]{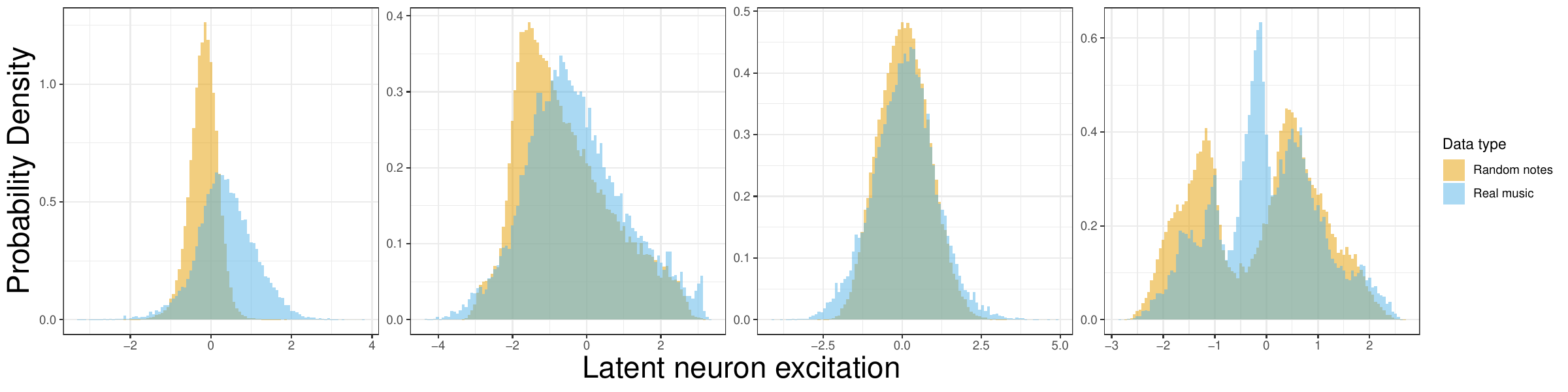}
\caption{\label{fig_real_vs_random_neuron_distros}Distributions of the excitation of the first four neurons in order of specificity for real music and random sequences.}
\end{figure}

In Figure~\ref{fig_real_vs_random_neuron_distros}, we show the histograms for the excitation of the four main neurons we identified above, in the case of the 50,000 melodies extracted from real music, and the 50,000 random note sequences we generated.

We see that the pitches of our random sequences differ from those of real music, as could be expected given the very broad distribution of pitches in our random sequences, whereas real melodies have narrower ranges of pitches.
As for rhythm on the other hand, the distribution of our random sequences of notes does not seem to differ that much from that of real music. This could be because we are only looking at 2 bars of music, admittedly a short sequence of notes most of the time, or because looking at one single latent dimension at a time is not enough to distinguish real music from random notes.

\subsection{Counting activated neurons}

\begin{figure}[h!]
\centering 
\includegraphics[width=1\textwidth,
]{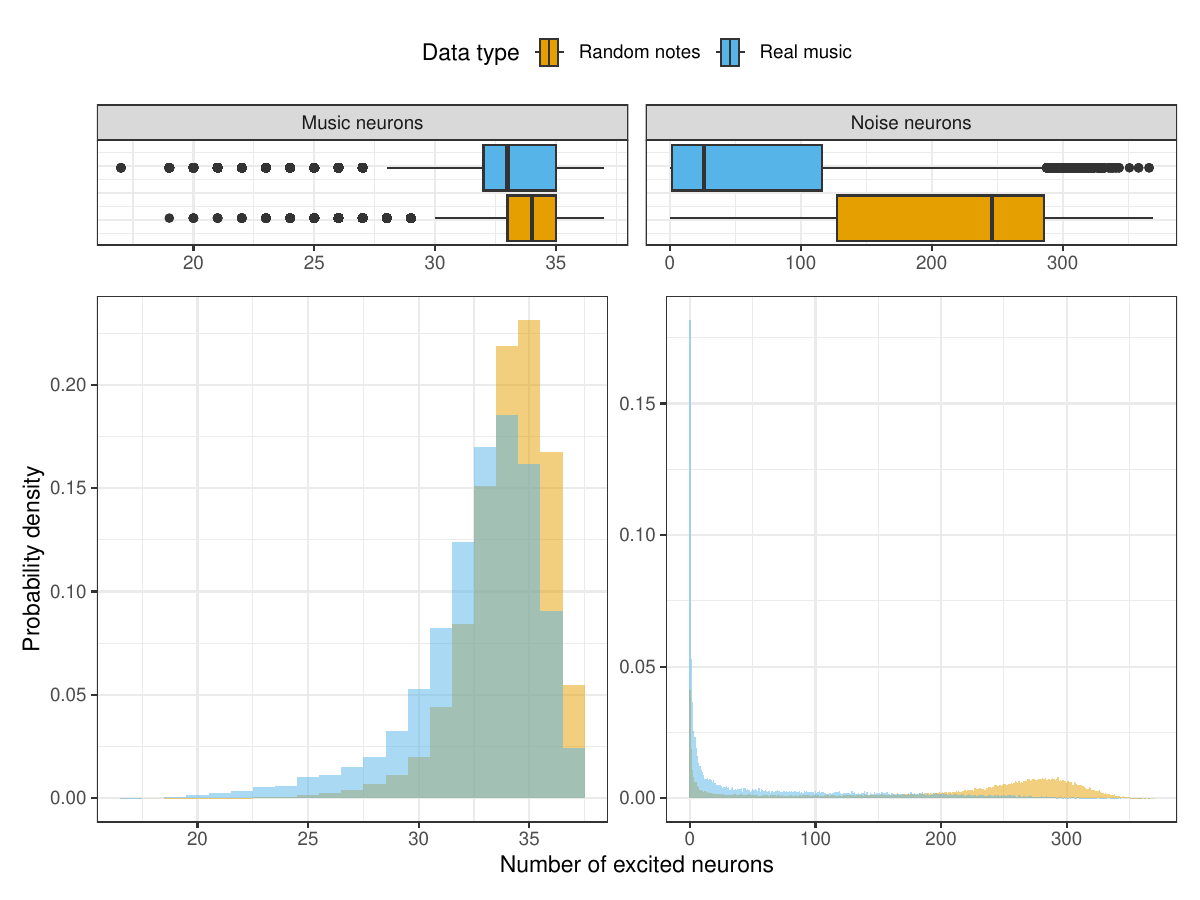}
\caption{\label{count_activations} Separate histograms of the number of latent music and noise neurons with (absolute value of) excitation larger than 0.1, for a random set of real songs and a set of random note sequences.}
\end{figure}

Instead of looking at the individual neurons, we count the number of latent dimensions for which the absolute value of the central value is larger than 0.1. For random note sequences that are not part of the data the VAE was trained on, one expects stronger excitations in the latent space. This is indeed the case, as shown in Figure~\ref{count_activations}: for the music neurons, both real music and random sequences excite nearly the same number of neurons: both distributions are peaked around 34 or 35 excited neurons out of 37 music neurons. As for noise neurons, real music tends to excite  fewer than 100 of them, while the distribution for random note sequences is bimodal, with most sequences exciting over a hundred noise neurons.

\section{Looking for melody (neurons)}
\label{16bar_section}

As we can see from Figure~\ref{fig_jsymb_latent_corrs_2bar}, we haven't found a neuron that can be conclusively said to encapsulate the melody information in the 2-bar case, at least not independently from rhythm: the second music neuron was correlated with many melody (M)  features, but even more so to rhythm (R) features. This could be due to the fact that 2 bars for music are not enough to give a strong melodic signal, and we therefore turn to the 16-bar case.

\subsection{Specifics of the 16-bar case}

\begin{figure}[h!]
\centering 
\includegraphics[width=1\textwidth,
]{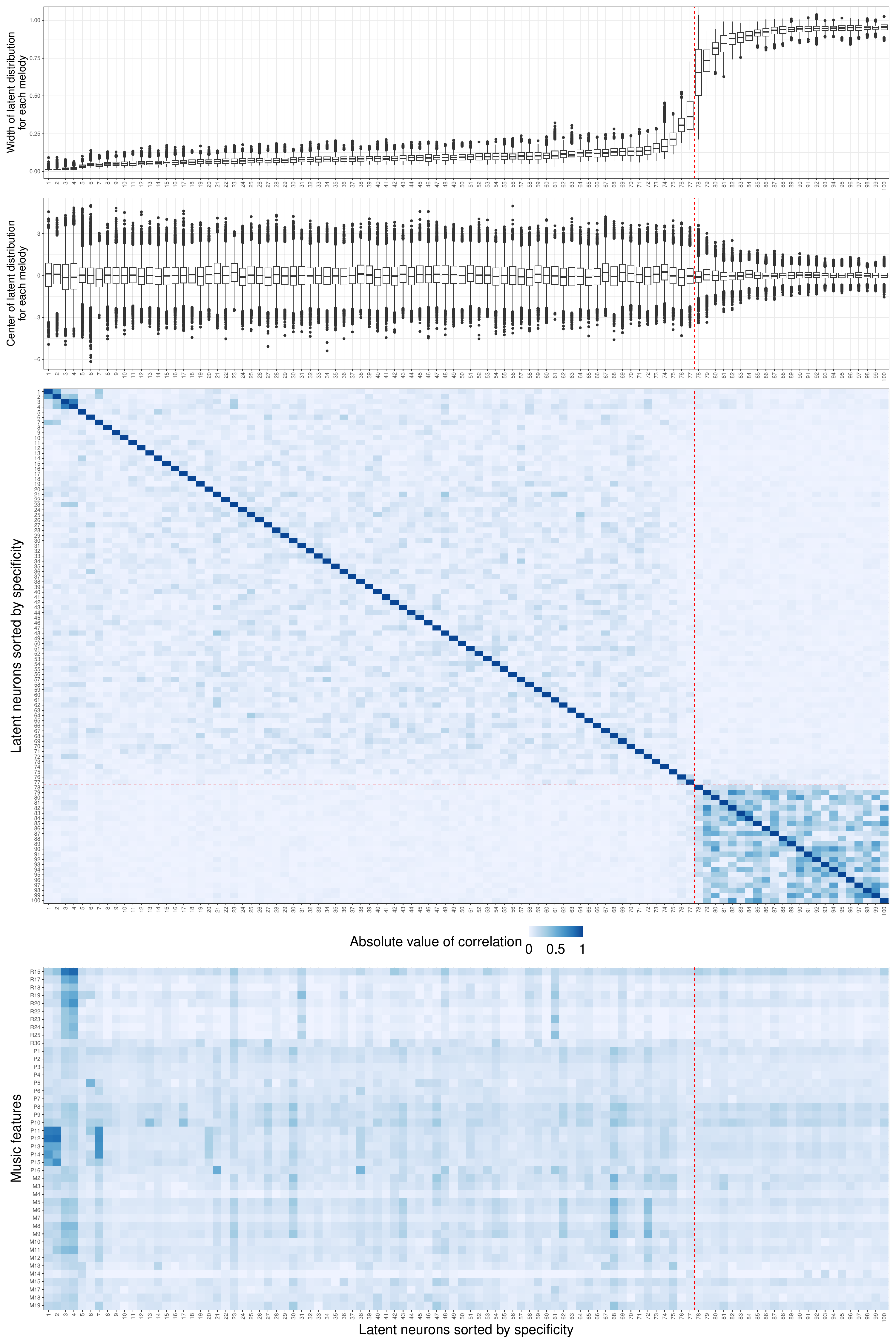}
\caption{\label{fig_latent_16bar} Same plots as above, but for the 16-bar model.}
\end{figure}

Figure~\ref{fig_latent_16bar} collects the information that was presented in the previous sections for the case of 2 bars of music, but applied two music sequences of 16 bars.
We find that the 16-bar case requires about double the number of music neurons than the 2-bar case: precisely 77 music neurons.

From this Figure, we can pick out music neurons 1 and 2, which are heavily correlated with pitch features, and music neurons 3 and 4, which are heavily correlated with rhythm features. These rhythm neurons are also heavily correlated with melody, but we are still looking for neurons that correlated more with melody features than rhythm or pitch. Possible candidates show up in Figure~\ref{fig_latent_16bar}, but much lower down the list of neurons, for instance neurons 23, 30, 43, 62, 68 and 72 are possible candidates. This might indicate that MusicVAE does not rely strongly on melody to organize its understanding of music, but that melody either appears as a consequence of the mode basic concepts of pitch and rhythm, or only plays a secondary role.



\section{Conclusion}

We have studied how a VAE trained on a million music tracks organizes its 512-dimensional latent space into hundreds of "noise neurons" that are barely used to encode music, and a few dozens of "music neurons" that actually encode musical information. It does make sense that the VAE only uses a fraction of its 512 dimensions for music: even the space of random note sequences would not require such a large latent space. In fact, we have found that random note sequences tend to excite 350 or fewer neurons (most of them noise neurons).

Returning to the case of real music, and in particular 2-bar melodies, we find that MusicVAE uses 37 music neurons to actually encode musical information, and of these, the first two in order of importance can be clearly identified as corresponding to pitch and rhythm.

Indeed, we show that several quantities defined in the literature to describe pitch are strongly correlated with the first music neuron, which non-linearly encodes this information into a single real variable. The same occurs for rhythm, but not for melody, which does not appear to be encoded independently from rhythm for such short music tracks.

Moving on to chunks of 16 bars of music, we see that MusicVAE uses 77 music neurons to encode music, but most pitch information is encoded non-linearly into the first two most important music neurons. The next two music neurons by order of importance encode several rhythm features. Dedicated neurons that only encode melody only show up much further in order of importance.

Whereas previous approaches have focused on enforcing a linear mapping of the human-defined quantities onto the latent space, we suggest that the non-linear change of representation is what allows the VAE to extract “principal coordinates" that diagonalize the problem, thereby simplifying and extending the human-defined variables.

\bmhead{Acknowledgments}


GB and JH acknowledge support from the Spanish grants  
PID2020-113334GB-I00 / AEI / 10.13039/501100011033 and CIPROM/2021/054
(Generalitat Valenciana). 
VS acknowledges support from the Generalitat
Valenciana PROMETEO/2021/083 and the Ministerio de Ciencia e
Innovacion PID2020-113644GB-I00. This project has received funding /support from the European Union’s Horizon 2020 research and innovation programme under the Marie Sk\l{}odowska-Curie grant agreement 860881-HIDDeN.


\bibliographystyle{plain} 

\end{document}